\documentclass[preprint]{aastex}

\pdfoutput=1

\slugcomment{Accepted for publication in \emph{ApJ Letters}}
\shorttitle{Rotation State of 103P from Radio Spectroscopy}
\shortauthors{Drahus et al.}

\begin{document}

\title{\Large Rotation State of Comet 103P/Hartley~2 from Radio Spectroscopy at
1~mm\altaffilmark{*}}

\author{Micha{\l} Drahus, David Jewitt, and Aur\'elie Guilbert-Lepoutre}
\affil{Department of Earth and Space Sciences, University of
California at Los Angeles,\\ Los Angeles, CA~90095}
\email{mdrahus@ucla.edu}

\author{Wac{\l}aw Waniak}
\affil{Astronomical Observatory, Jagiellonian University, Krak\'ow,
Poland}

\author{James Hoge}
\affil{Joint Astronomy Centre, Hilo, HI~96720}

\author{Dariusz C. Lis}
\affil{Division of Physics, Mathematics, and Astronomy, California
Institute of Technology,\\ Pasadena, CA~91125}

\author{Hiroshige Yoshida and Ruisheng Peng}
\affil{Caltech Submillimeter Observatory, Hilo, HI~96720}

\and

\author{Albrecht Sievers}
\affil{Instituto de Radio Astronom\`ia Milim\'etrica, Granada,
Spain}

\altaffiltext{*}{Based on observations carried out with the IRAM
\mbox{30-m}, JCMT \mbox{15-m}, and CSO \mbox{10.4-m} telescopes.
IRAM is supported by INSU/CNRS (France), MPG (Germany), and IGN
(Spain). JCMT is operated by Joint Astronomy Centre and supported by
STFC (UK), NRC (Canada), and NWO (Netherlands). CSO is operated by
Caltech and supported through NSF grant \mbox{AST-0540882} (USA).}

\begin{abstract}
The nuclei of active comets emit molecules anisotropically from
discrete vents. As the nucleus rotates, we expect to observe
periodic variability in the molecular emission line profiles, which
can be studied through mm/submm spectroscopy. Using this technique
we investigated the HCN atmosphere of comet \mbox{103P/Hartley 2},
the target of NASA's EPOXI mission, which had an exceptionally
favorable apparition in late 2010. We detected short-term evolution
of the spectral line profile, which was stimulated by the nucleus
rotation, and which provides evidence for rapid deceleration and
excitation of the rotation state. The measured rate of change in the
rotation period is $+1.00 \pm 0.15$ min per day and the period
itself is $18.32 \pm 0.03$ hr, both applicable at the epoch of the
EPOXI encounter. Surprisingly, the spin-down efficiency is lower by
two orders of magnitude than the measurement in comet
\mbox{9P/Tempel 1} and the best theoretical prediction. This secures
rotational stability of the comet's nucleus during the next few
returns, although we anticipate a catastrophic disruption from
spin-up as its ultimate fate.
\end{abstract}


\section{Introduction}

\mbox{103P/Hartley 2} (hereafter 103P) is a Jupiter Family comet
(JFC), which currently has a \mbox{6.47-year} orbital period and
perihelion at 1.06~AU. On UT~2010 Oct.~20.7 it reached the minimum
geocentric distance of only 0.12~AU, making by far the closest
approach to the Earth since its discovery \citep{Har86}, and
becoming a naked-eye object. Shortly after, on UT~2010 Nov.~4.6, the
comet was visited by NASA's EPOXI spacecraft, which approached the
nucleus within 700~km and provided detailed images and spectra
\citep{AHe11}. Both the ground-based data, obtained at the unusually
favorable geometry, and the unique observations carried out by the
spacecraft, create an exceptional platform for new groundbreaking
investigations.

Taking advantage of this unique opportunity we addressed one of the
burning problems of planetary science, which is the characteristic
lifetime of active cometary nuclei. They can be dynamically ejected
out of the sublimation zone, or decay through collision, tidal
disruption, devolatilization, deactivation, disintegration, and --
the least understood -- spontaneous fragmentation \citep{Jew92}. Of
several scenarios proposed to explain the last process, a notable
one is \emph{rotational break-up}, which must occur for spinning
bodies when the centripetal force surpasses self-gravity and
material strength \citep[e.g.][]{Dav01}. Interestingly, models show
that torques exerted by gas jets can significantly increase or
decrease the rotation rate during a single orbit
\citep[e.g.][]{Gut02}, although the magnitude strongly depends on
the nucleus size and activity, and the poorly understood effective
moment arm. The first unambiguous measurement of this effect was
obtained only recently for comet \mbox{9P/Tempel 1}, which was found
to be slowly spinning up \citep{Bel07,Bel11}. Clearly, further
observational studies of the dynamics of cometary rotation are
essential to establish the nucleus lifetime with respect to
rotational break-up.

Since the influence of jets must be strongest for small, elongated,
and active comets \citep[e.g.][]{Sam86,Jew99,Dra06}, the nucleus of
103P is an excellent test body: it is one of the smallest known JFCs
(equivalent radius $\sim600$~m), it is elongated and active
\citep{Lis09,AHe11}, and hence was predicted to show measurable
period changes during the last return \citep{Dra09,Lis09}. In
general, however, it is very difficult to investigate the rotation
of an active nucleus; especially of a very small one and from the
ground -- securing simultaneously the necessary measurement
precision. Our earlier studies show that perhaps the best
possibilities are offered by mm/submm spectroscopy \citep{Dra10}.
When the molecules are emitted from discrete vents, or the body is
irregular, the diurnal rotation modulates the emission line
profiles, which can be conveniently studied in a velocity-resolved
spectral time series. The effect will be best visible when every
single spectrum results from the molecules released at similar
nucleus rotation phases, which is satisfied when the molecules leave
the projected telescope beam on a timescale much shorter than the
body rotation period. The best candidates are hence nearby comets
rotating slowly, and observed ideally in small beams using short
integration times with the largest antennas. Consequently, we
identified 103P as a terrific target for large ground-based mm/submm
telescopes \citep{Dra09}, owing to the unusually small geocentric
distance, localized activity, and more than sufficiently slow
rotation (the last two confirmed after the fact).

\section{Observations}

Between early September and mid December 2010 we used all large
single-dish ground-based mm/submm facilities operating around one
millimeter and offering open time, to carry out spectral monitoring
of HCN. This molecule is a particularly good tracer of the cometary
rotation \citep{Dra10} thanks to its well-established origin
directly from the nucleus and because it has by far the brightest
emission lines in the one-millimeter atmospheric windows. Our
campaign was the first at these wavelengths specifically designed to
investigate the rotation state of a comet, and thus provided an
unprecedentedly rich and dense velocity-resolved spectral time
series of a cometary parent molecule.

\begin{figure}
\epsscale{.50} \plotone{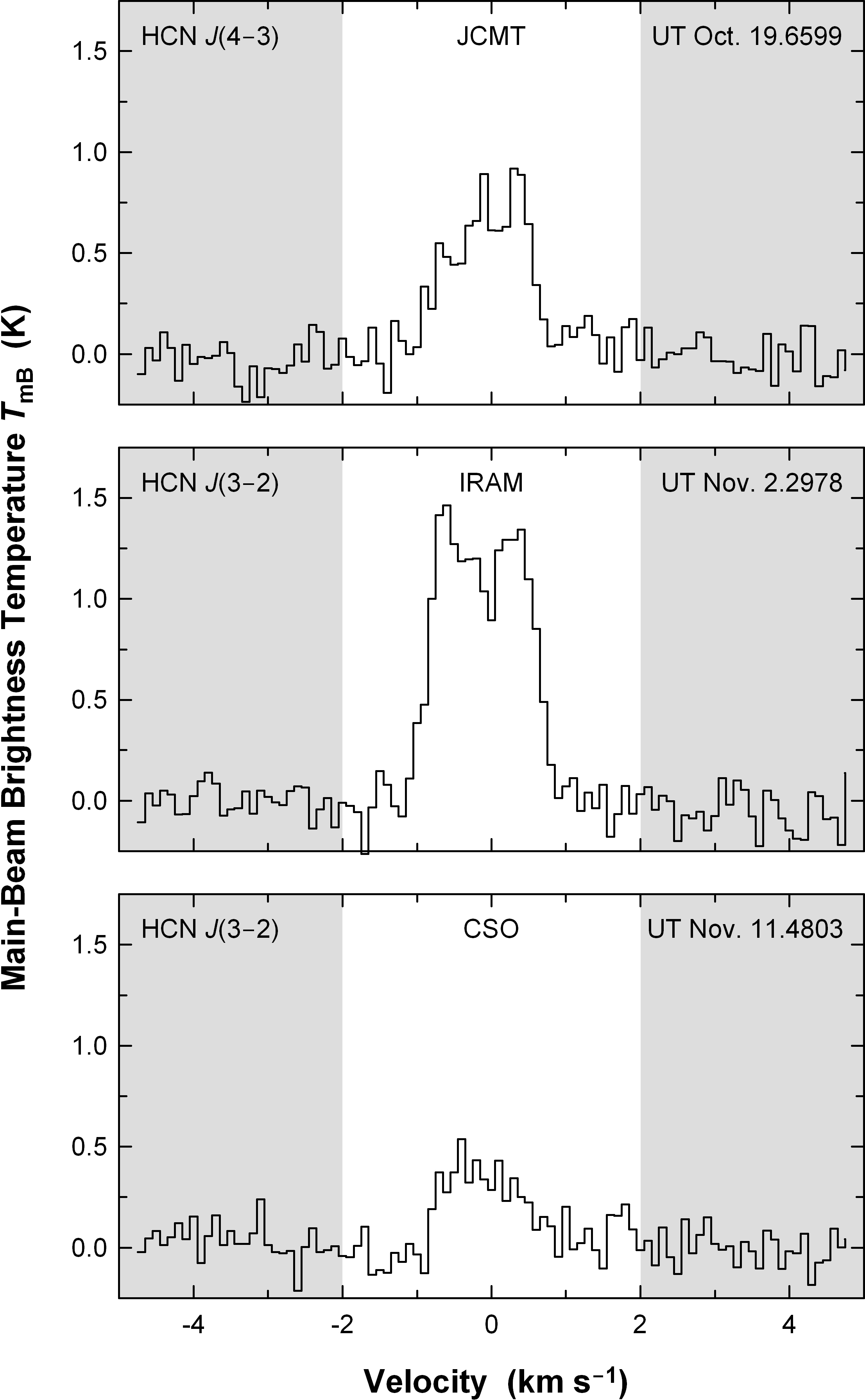} \caption{Example of HR
spectra. The unshaded region was used for calculations.\label{fig1}}
\end{figure}

In this work we use 438 spectra of the \mbox{\textit{J}(3--2)} and
\mbox{\textit{J}(4--3)} rotational transitions (Fig.~\ref{fig1}),
collected with IRAM, JCMT, and CSO on 20 nights between UT~2010
Sep.~29.3 and Dec.~15.6 (Table~\ref{tab1}). The spectra were
calibrated in the standard manner; each covers 10 to 15~min, and was
analyzed at spectral resolutions of 0.10~km~s$^{-1}$ (high
resolution; HR) and 0.25~km~s$^{-1}$ (low resolution; LR). Details
of the observations and reductions will be presented in a subsequent
paper.

\section{Analysis}

The HR spectra were first parameterized in terms of the line area
and center velocity and then converted to physical quantities: HCN
production rate $Q$ and median radial HCN gas-flow velocity
$v_\mathrm{rad}$. All these steps were performed following our
standard methods and procedures \citep{Dra09,Dra10}, and will be
described in a subsequent paper. In this work we only highlight our
approach to estimate errors (Appendix~\ref{Errors}).

\begin{figure}
\epsscale{.50} \plotone{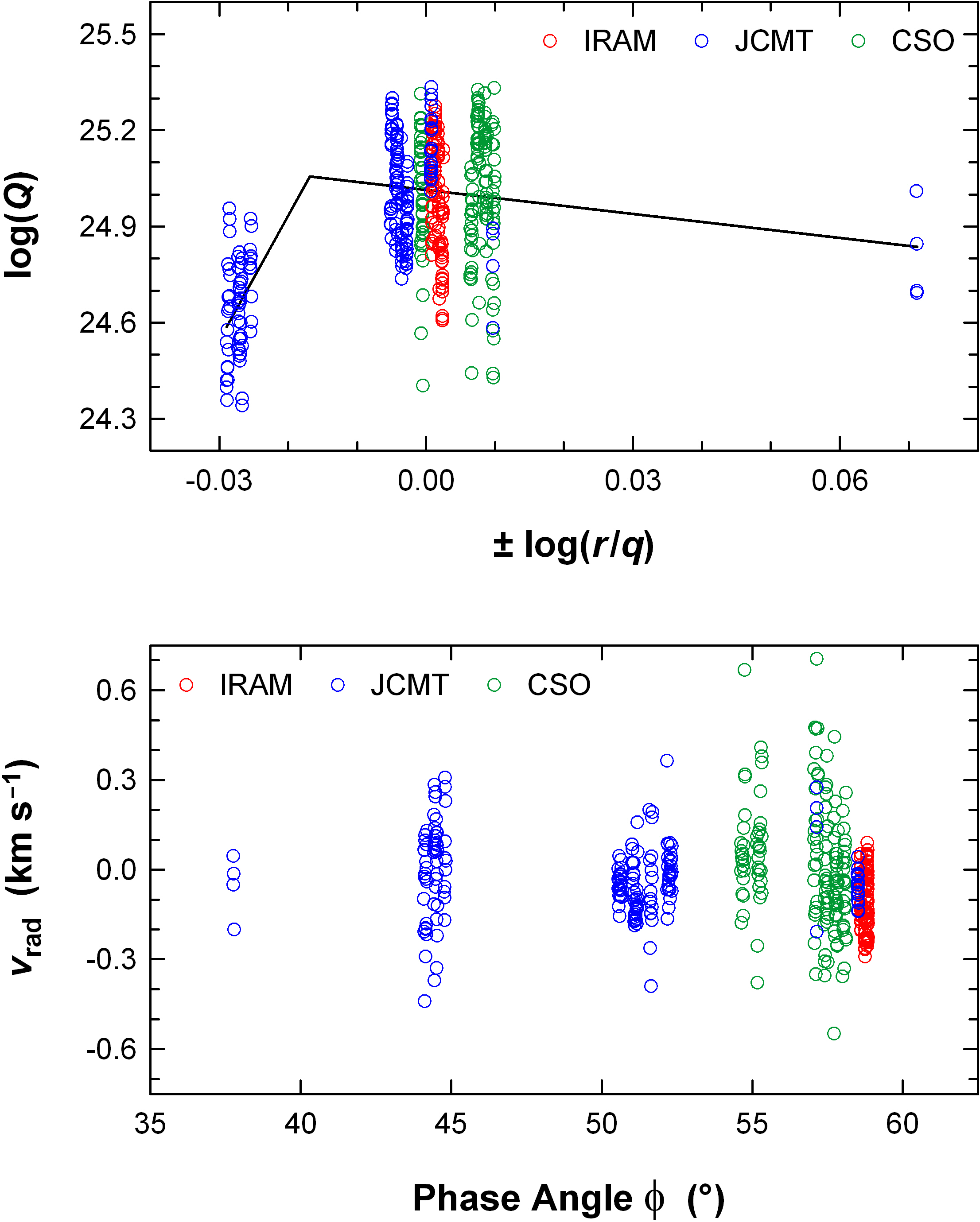} \caption{Orbital trends in
the data. \emph{Top panel:} Dependence of the production rate $Q$ on
heliocentric distance $r$ (normalized to the perihelion distance
$q$) in a log--log space. Negative $\log(r/q)$ indicate positions
before perihelion. The solid line shows a semi-empirical fit that
was used to remove the heliocentric trend and calculate the
production-rate deviations $\Delta\log(Q)$. \emph{Bottom panel:}
Behavior of the median radial velocity $v_\mathrm{rad}$ with phase
angle $\phi$. \emph{Note:} The EPOXI encounter occurred at
$\log(r/q) = 0.002$ and $\phi = 58.8\degr$.\label{fig2}}
\end{figure}

Although unavoidably affected by model simplifications, the physical
quantities are naturally free (to the first order) of trivial
instrumental, topocentric, and transition-specific differences,
hence making the data set fairly homogeneous and suitable for the
investigation of variability. Figure~\ref{fig2} shows the evolution
of $Q$ with heliocentric distance $r$, and displays $v_\mathrm{rad}$
against phase angle $\phi$.

\begin{figure}
\epsscale{.50} \plotone{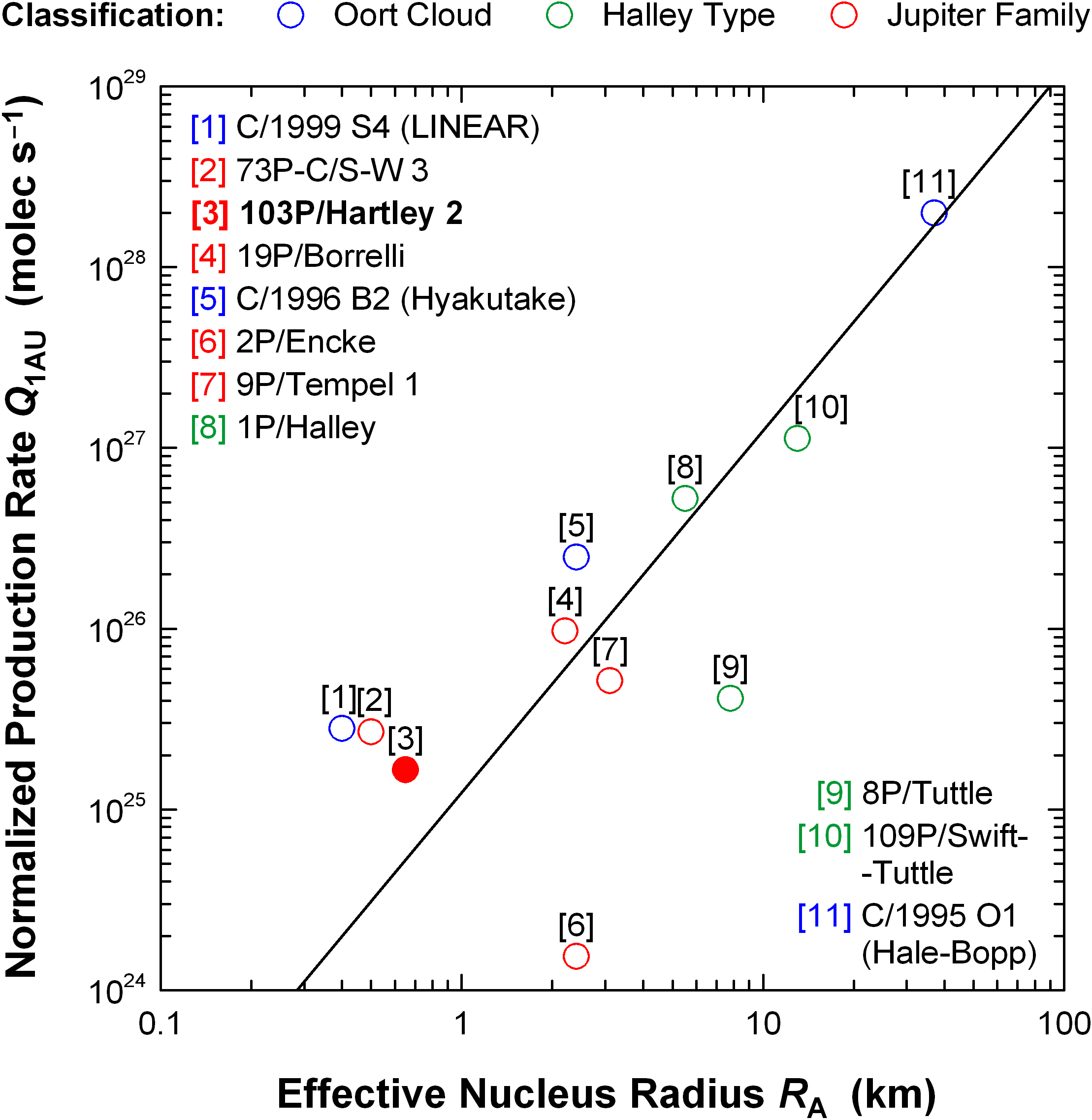} \caption{Compilation of the
normalized HCN production rates $Q_\mathrm{1AU}$ for comets with
known area-equivalent nucleus radius $R_\mathrm{A}$, presented in a
log--log space \citep[see][and references therein]{Dra10}. The
production rates were normalized to the common heliocentric distance
$r = 1$~AU using the canonical $Q(r) \sim r^{-4}$ whenever needed.
The solid line is a square function fitted in a log--log space. It
represents a typical (constant) active fraction of the
nucleus.\label{fig3}}
\end{figure}

The first trend in Fig.~\ref{fig2} was measured for several comets
before \citep[e.g.][]{Biv02a,Dra10}. We see an increase in the
production rate as 103P approaches perihelion and a decrease
afterwards, with superimposed short-term variability responsible for
the excessive scatter. The maximum mean-diurnal level of HCN is
about $1 \times 10^{25}$ molec~s$^{-1}$, which seems high for a JFC
but not abnormal for a comet (Fig.~\ref{fig3}). The ratio of HCN to
H$_2$O \citep{Biv10,AHe11} at the level of 0.1\% is also typical of
comets \citep{Biv02b,Boc04}. After subtracting the heliocentric
trend we derived production-rate deviations $\Delta\log(Q)$ about
the mean-diurnal level.

The expected correlation of $v_\mathrm{rad}$ with $\phi$ has been
observationally confirmed only recently \citep{Dra10}; however, in
our data we do not find any trend (Fig.~\ref{fig2}), because of the
highly limited phase-angle coverage, short-term variations, and
observing noise.

\section{Nucleus Rotation}

While the short-term variations could be sporadic \citep{AHe05}, it
is also entirely possible that they were stimulated by the nucleus
rotation. We pursued this issue using an innovative technique: the
\emph{Dynamized Structural Periodicity Analysis} (DSPA), which we
created specifically for this purpose (Appendix~\ref{DSPA}). The
method was developed from our \emph{Dynamized Periodicity Analysis}
(DPA), which -- utilizing classical algorithms for a time series of
signals -- introduces a dynamical formula for the body rotation and
rigorous weighting of the input data \citep{Dra06}. DPA determines
the rotation frequency simultaneously with its time-derivative from
the entire data set, taking advantage of the maximum available time
base, and therefore is far superior compared to separate classical
(constant) period determinations from smaller subsets. Recently DPA
demonstrated excellent sensitivity in providing important evidence
for the slow spin-up of comet \mbox{9P/Tempel 1}
\citep{Bel07,Bel11}. DSPA goes further, and analyzes periodicity not
only in the level, but in the complete structure of the signal in
the input data, properly combining the information from independent
data channels. At the same time it keeps all the properties of DPA,
and for a single-channel time series fully reduces to the latter.
The technique does not rely on any variability models, neither
versus time nor across the channels. Although DSPA allows for any
a-priori law controlling the effective torque, in this analysis we
assumed it was constant.

Using the derived production-rate variations $\Delta\log(Q)$, we
normalized the original spectra to eliminate all the trivial effects
and remove the physical heliocentric trend (Fig.~\ref{fig2}). In
order to secure an affordable computation time and reasonable S/N in
the individual data channels, DSPA was applied to the normalized LR
spectra. We used 17 channels from the velocity interval between -2
and +2~km~s$^{-1}$, which is more than enough to adequately sample
the line profile. They were analyzed with two well-established
kernels: weighted version of \emph{Phase Dispersion Minimization}
(PDM) and weighted \emph{Harmonic Fit} (HF) \citep{Dra06}.

\begin{figure}
\epsscale{0.86} \plotone{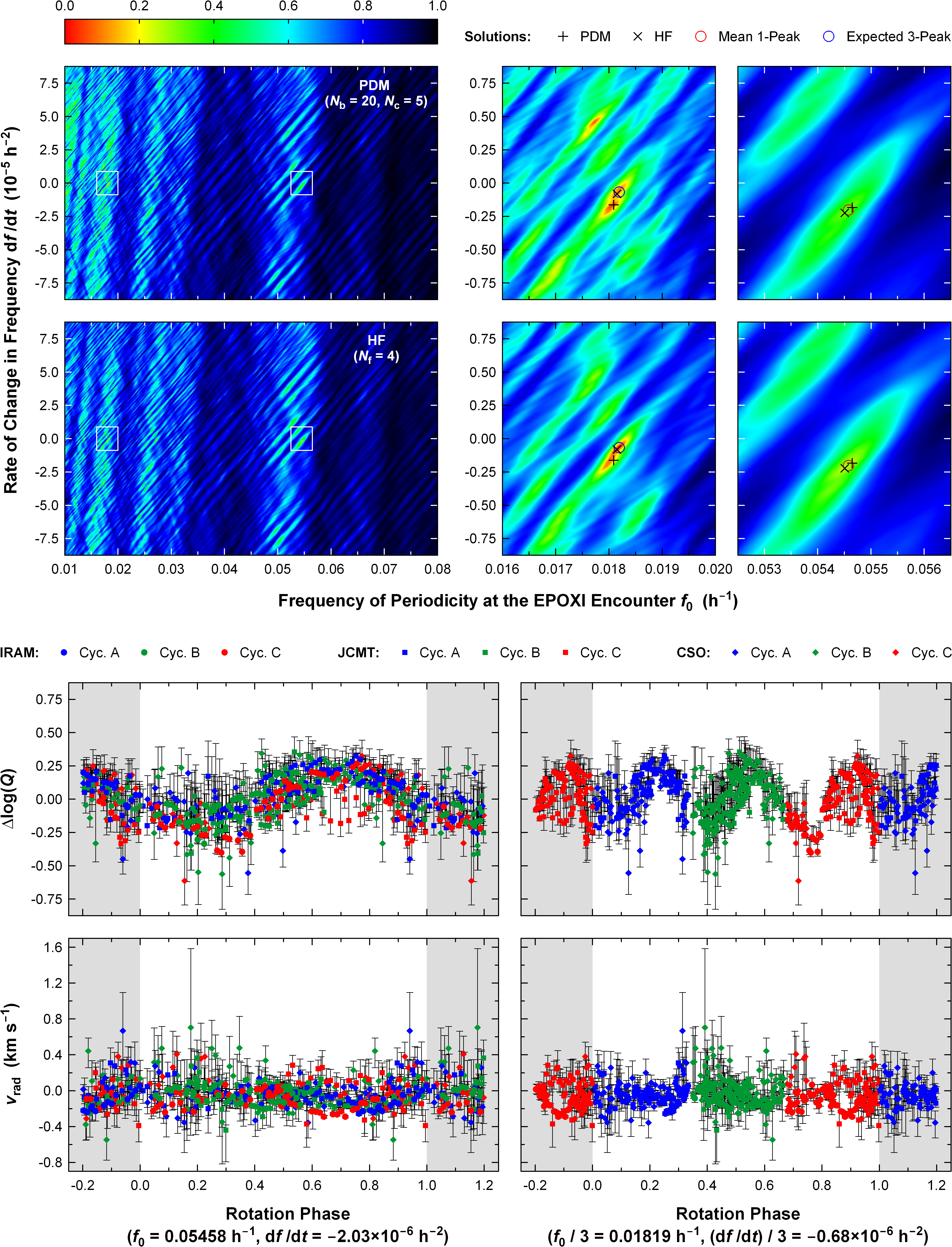} \caption{Results of DSPA.
\emph{Upper panels:} Dynamical periodograms for the regions of
investigated space. Linear color scale denotes $\theta_\mathrm{disp}
=
\log(\eta(\overline{\theta}-\overline{\theta}_\mathrm{min})/(\overline{\theta}_\mathrm{max}-\overline{\theta}_\mathrm{min})+10-\eta)$,
where $\eta = 9$ and function
$\overline{\theta}=\overline{\theta}(f,\mathrm{d}\!f/\mathrm{d}t)$
is defined in Appendix~\ref{DSPA}. \emph{Lower panels:} Rotation
phase profiles for $\Delta\log(Q)$ and $v_\mathrm{rad}$ for two
periodicity solutions (see text). The shaded regions indicate
duplications of the phase space. The EPOXI encounter occurred during
\emph{Cycle~B} at the rotation phase of 0.5.\label{fig4}}
\end{figure}

Figure~\ref{fig4} presents \emph{dynamical periodograms} resulting
from the analysis. The best single-peak solution is found for the
rotation frequency $f_0 = 0.0546 \pm 0.0001$ h$^{-1}$ (given at the
epoch of the EPOXI encounter; indicated by index zero) and frequency
time-derivative $\mathrm{d}\!f/\mathrm{d}t = (-2.0 \pm 0.3) \times
10^{-6}$ h$^{-2}$. It is equivalent to the period of $18.32 \pm
0.03$ h increasing at the relative rate of $0.07\% \pm 0.01\%$
(e.g., $1.00 \pm 0.15$ min/day, or $0.75 \pm 0.10$ min/cycle), both
applicable at the EPOXI encounter. The result is insensitive to the
uncertainties resulting from the earlier reductions -- including the
removal of the heliocentric trend, as the data most affected enter
the analysis with very low relative statistical weights.

We note that the global solution is very close to the third multiple
of the above single-peak solution, which means that the pattern of
mass loss repeats significantly better every three rotation cycles.
We explain this behavior by excitation of the rotation state: every
three cycles of the fundamental periodicity the independent rotation
modes coincide at similar phases and then the \mbox{\emph{3-cycle}}
repeats. But the resonance is not perfect. The spectra from the same
rotation phases and corresponding cycles still show differences in
the line profiles (e.g., Fig.~\ref{fig1} shows spectra from phase
0.5 and \emph{Cycle~B} [cf. Fig.~\ref{fig4}], corresponding to the
moment of the EPOXI encounter). We attribute this effect primarily
to small differences in the instantaneous nucleus orientation, which
cause differences in the direction of gas flow. Interestingly, the
larger nucleus of comet \mbox{9P/Tempel 1} was found close to a
fully-relaxed rotation state \citep{Bel11}, while the even bigger
nucleus of \mbox{1P/Halley} was in a significantly excited state
\citep{Bel91}.

Figure~\ref{fig4} also shows the production-rate variations
$\Delta\log(Q)$ and median radial gas-flow velocities
$v_\mathrm{rad}$ phased according to the best single-peak solution
and its third multiple. The line variability comes primarily from
the diurnal changes in the production rate, which follow a nearly
sinusoidal trend. The single-peak solution is consistent with a
single strong active area being the primary source of HCN. As the
EPOXI encounter occurred at the rotation phase of 0.5, when the HCN
production rate was shortly before the diurnal maximum, we suppose
that this area coincides with the strongest jet observed by the
spacecraft \citep{AHe11}. This also implies that the higher-order
multi-peak solutions, including the triple-peak third multiple
indicated by the global minimum, cannot be plausibly associated with
the fundamental periodicity.

\section{Discussion and Conclusions}

Our best solution indicates a rapid spin-down of 103P's nucleus.
Spin-down at a comparable rate was suggested earlier for comet
\mbox{C/2001 K5} \citep{Dra06}, but the only robust measurement,
obtained for \mbox{9P/Tempel 1}, indicates spin-up at a
\mbox{20-times} smaller rate \citep{Bel07,Bel11}. Dynamics of the
rotation rate can be interpreted in terms of a simple model
\citep[e.g.][]{Sam86,Jew99,Dra06}, where the rate of change in
frequency $\mathrm{d}\!f/\mathrm{d}t$ is given by:
\begin{equation}\label{eq1}
\mathrm{d}\!f/\mathrm{d}t =
\frac{15}{16\pi^2}\frac{v_\mathrm{s}\,Q_\mathrm{tot}}{R_\mathrm{V}^4\,
\rho}\kappa,
\end{equation}
where $v_\mathrm{s}$ is the gas sublimation velocity,
$Q_\mathrm{tot}$ is the (mean-diurnal) production rate of all
gaseous species, $R_\mathrm{V}$ is the volume-equivalent nucleus
radius, $\rho$ is the nucleus bulk density, and $\kappa$ is the
dimensionless effective moment arm which is a measure of
acceleration efficiency; $\kappa$ is negative for spin-down and
positive for spin-up, and may exceed unity for non-spherical bodies
\citep{Dra06}. Substituting the measured $\mathrm{d}\!f/\mathrm{d}t
= -2.0 \times 10^{-6}$ h$^{-2}$, and $Q_\mathrm{tot} =
Q_\mathrm{H_2O} + Q_\mathrm{CO_2} = 450$ kg~s$^{-1}$ \citep{AHe11},
$v_\mathrm{s} = 400$ m~s$^{-1}$ \citep{Com04}, $\rho = 400$
kg~m$^{-3}$ \citep{AHe05}, and $R_\mathrm{V} = 580$~m
\citep{Lis09,AHe11}, we find $\kappa = -0.04\%$. The same equation
yields $\kappa = +4\%$ for the properties of comet \mbox{9P/Tempel
1} ($R_\mathrm{V} = 3000$ m [\mbox{A'Hearn et al.} 2005],
$Q_\mathrm{tot} = Q_\mathrm{H_2O} + Q_\mathrm{CO_2} = 160$
kg~s$^{-1}$ [\mbox{Feaga et al.} 2007], and
$\mathrm{d}\!f/\mathrm{d}t = +9.7 \times 10^{-8}$ h$^{-2}$
[\mbox{Belton et al.} 2011]; otherwise the same as above), which is
consistent with a model prediction $|\kappa| \sim 5\%$
\citep{Jew99}. Even though $\kappa$ in 103P is two orders of
magnitude lower compared to 9P, the comet changes its rotation
frequency a factor of 20 faster. This is understandable from
Eq.~(\ref{eq1}), since $\mathrm{d}\!f/\mathrm{d}t$ has a strong
dependence on the nucleus size, and the nucleus of 103P is tiny
compared to 9P. Moreover, 103P is more active per unit surface area
(Fig.~\ref{fig3}) and approaches the Sun closer, therefore it has a
higher total sublimation rate at perihelion, in spite of the much
smaller total area.

If the pattern of mass loss remains stable, the above result lets us
crudely speculate about the rotational past and future of 103P on
the current orbit. We estimate the orbit-integrated frequency change
at the level of 0.012~h$^{-1}$ and assume the critical rotation
period of approximately 3~h \citep{Dav01}. This allows for no more
than $\sim20$ previous revolutions. At the present rate, the nucleus
will stop rotating during the fourth or fifth return from now
(perihelion in 2036 or 2043). If it survives the increased thermal
stress, it should start spinning up in the opposite direction, and
reach the rotational disruption limit $\sim25$ orbits later
\mbox{(year $\sim2200$)}. We note that rotational break-up has by
far the shortest timescale determining the lifetime of such a small
object \citep{Jew92,Jew99,Lis09}. Moreover, bearing in mind the
location of the main active area on the nucleus \citep{AHe11}, even
a small change in the jet configuration may dramatically increase
the effective moment arm -- bringing 103P to the end of its life
within one or few orbits.

We expect that small, young, and naturally volatile-rich comets,
having typical effective moment arms of the order of 5\%, routinely
experience rotational instability and fragmentation right after
being injected to the sublimation zone yet before being discovered.
In this way the observed flat size distribution of comet nuclei
\citep[e.g.][]{Sno11} can be naturally explained. With its tiny and
active nucleus, 103P appears as a lucky survivor, protected for some
limited time from rotational disruption by the remarkably low
factor~$\kappa$.

\acknowledgments

We thank M.~Poli\'nska for providing resources and the telescope
staffs for their excellent work. This project was supported by NASA
through the Planetary Astronomy Program grant to D.~Jewitt.

{\it Facilities:} \facility{IRAM 30-m (EMIR)}, \facility{JCMT
(HARP)}, \facility{CSO (Z-Rex)}.

\appendix

\section{Error Evaluation}\label{Errors}

We have examined all error sources and concluded that the signal
uncertainty in the individual spectral channels is dominated by
observing noise and imperfect pointing.

Observing noise was estimated from signal variations about the
baseline assuming a normal distribution. Since noise is nearly
entirely controlled by the sky and receiver temperatures (surpassing
the comet antenna temperature by about two orders of magnitude), we
assumed the same noise level inside the line window. Noise
dependence on frequency is negligible in this extremely narrow
spectral range.

The influence of pointing errors was estimated from a simple model,
which we have developed for this purpose from our earlier
constructions \citep{Dra10}. For each telescope axis we assume a
normally-distributed pointing error with a standard deviation of
$2\arcsec$ for IRAM and JCMT, and $3\arcsec$ for CSO; it is easy to
show that this is equal to the most probable total pointing offset.
We also adopt a simple steady-state isotropic coma model with
infinite molecular lifetime and brightness proportional to the
number of molecules. Under these assumptions we calculated
Probability Density Functions (PDFs) of signal loss for each
telescope, and neglected other (less significant) effects from
imperfect pointing.

To calculate the errors of all the parameters of interest we
followed our Monte-Carlo approach \citep{Dra06,Dra10}, and generated
500 simulations using the derived PDFs. A starting point for each
simulation was an original spectrum corrected for the most probable
signal loss from pointing. To each spectral channel we added a
random realization of noise and then scaled the complete spectrum
according to a random realization of signal loss from pointing.
Then, by analyzing the parameters' variations about their original
values, we derived their standard errors.

The signal errors in the individual spectral channels feature
several interesting properties. With signal approaching zero they
asymptotically approach the noise limit, and oppositely, they
approach the pointing limit with the signal increasing to infinity.
The pointing component is proportional to signal whereas the noise
component does not depend on signal. While the noise component can
be controlled (reduced) by spectral or temporal binning, the
pointing component cannot, because it affects all the channels in
one spectrum in the same way, and similarly several spectra taken
within short time (the latter not being taken into account by our
simple model). For this reason the pointing component should not be
included when analyzing signal variability across the channels.
Moreover, the error from pointing is nearly one-sided (positive)
while the one from noise is symmetric.

Because several of the assumptions of our model may not be well
satisfied, the provided uncertainties should be considered as crude
estimates. Nevertheless, to our best knowledge it is the first
approach to quantify how the imperfect pointing of radio telescopes
affects the observed spectra.

\section{Dynamized Structural Periodicity Analysis}\label{DSPA}

The DSPA method is applied to a time series of $M$ measurements.
Quality of data phasing is measured by a function $\theta$, which is
a variance ratio of the phased and unphased data. DSPA calculates
functions $\theta_i(f,\mathrm{d}\!f/\mathrm{d}t)$ of DPA
\citep{Dra06} for every data channel $i$, and averages the results
from the considered $N$ channels to provide
$\overline{\theta}(f,\mathrm{d}\!f/\mathrm{d}t)$:
\begin{equation}
\overline{\theta} = \frac{\sum\limits_{i=1}^N
\theta_i\,W_i}{\sum\limits_{i=1}^N W_i},
\end{equation}
using weights $W_i$ defined as square-mean signal over mean-square
error:
\begin{equation}
W_i =
\frac{\overline{S_i}^2}{\overline{\sigma_i^2}}=\Bigg(\frac{\sum\limits_{j=1}^M
S_{ij}\,w_{ij}}{\sum\limits_{j=1}^M w_{ij}}\Bigg)^2\,
\frac{M}{\sum\limits_{j=1}^M \sigma_{ij}^2},
\end{equation}
where $S_{ij}$ is the signal in $i$-th data channel and $j$-th time
moment, $\sigma_{ij}$ is its standard error, and $w_{ij} =
\sigma_{ij}^{-2}$. Note that both functions $\theta_i$ used by DPA,
i.e. the weighted version of PDM and the weighted HF, use the same
weights $w_{ij}$ as above. Since our signal errors are asymmetric,
we used their mean-square values for this purpose.

Weighting with $W_i$ implemented in DSPA naturally prefers channels
with strong signal, making the method insensitive, from the
theoretical point of view, to the selection of line window (as long
as the line is fully inside); channels having on average zero signal
contribute to $\overline{\theta}$ with zero weights. Moreover, it
prefers channels with small uncertainties, which is particularly
important in this application since imperfect pointing
differentiates signal error across the channels; for identical
uncertainties in every channel this term cancels out. It is
therefore clear that in our case the effective channel weight is
established by the fine balance between channel signal and noise --
as the error from pointing is proportional to the signal.

\clearpage

\begin{deluxetable}{lcrrrrrrr}
\tabletypesize{\scriptsize} \tablecaption{Journal of
observations\label{tab1}} \tablewidth{0pt} \tablehead{ \colhead{UT
Date 2010} & \colhead{Telescope} & \colhead{Cover.\tablenotemark{a}}
& \colhead{Num.\tablenotemark{b}} & \colhead{$r$\tablenotemark{c}} &
\colhead{$\Delta$\tablenotemark{d}} &
\colhead{$\phi$\tablenotemark{e}} & \colhead{Beam\tablenotemark{f}} & \colhead{Escape\tablenotemark{g}} \\
\colhead{(mid time)} & \colhead{} & \colhead{(h)} & \colhead{} &
\colhead{(AU)} & \colhead{(AU)} & \colhead{(\degr)} & \colhead{(km)}
& \colhead{(min)}} \startdata
Sep. \hspace{0.058cm} 29.4795  &  JCMT  &   6.6  &  20  &  1.1310  &  0.1926  &  44.1  &   958  &  20.0  \\
Sep. \hspace{0.058cm} 30.4229  &  JCMT  &   8.1  &  28  &  1.1265  &  0.1873  &  44.5  &   932  &  19.4  \\
Oct. \hspace{0.190cm}  1.2865  &  JCMT  &   2.8  &  11  &  1.1225  &  0.1826  &  44.8  &   908  &  18.9  \\
Oct. \hspace{0.042cm} 16.6228  &  JCMT  &   6.5  &  21  &  1.0710  &  0.1241  &  50.6  &   617  &  12.9  \\
Oct. \hspace{0.042cm} 17.5706  &  JCMT  &  10.0  &  33  &  1.0691  &  0.1228  &  51.1  &   611  &  12.7  \\
Oct. \hspace{0.042cm} 18.5236  &  JCMT  &   5.0  &  16  &  1.0673  &  0.1218  &  51.6  &   606  &  12.6  \\
Oct. \hspace{0.042cm} 19.5997  &  JCMT  &   8.3  &  27  &  1.0655  &  0.1211  &  52.3  &   602  &  12.5  \\
Oct. \hspace{0.042cm} 23.6593  &   CSO  &   6.7  &  20  &  1.0606  &  0.1225  &  54.7  &  1249  &  26.0  \\
Oct. \hspace{0.042cm} 24.5773  &   CSO  &   8.3  &  24  &  1.0599  &  0.1236  &  55.2  &  1260  &  26.3  \\
Nov. \hspace{0.150cm}  1.6117  &  JCMT  &   6.1  &  20  &  1.0604  &  0.1447  &  58.5  &   720  &  15.0  \\
Nov. \hspace{0.150cm}  2.2058  &  IRAM  &   7.9  &  24  &  1.0609  &  0.1469  &  58.6  &   470  &   9.8  \\
Nov. \hspace{0.150cm}  3.1957  &  IRAM  &   8.5  &  26  &  1.0619  &  0.1506  &  58.7  &   482  &  10.1  \\
Nov. \hspace{0.150cm}  4.1974  &  IRAM  &   8.4  &  25  &  1.0631  &  0.1546  &  58.8  &   495  &  10.3  \\
Nov. \hspace{0.150cm}  5.1820  &  IRAM  &   8.7  &  28  &  1.0644  &  0.1586  &  58.8  &   508  &  10.6  \\
Nov. \hspace{0.000cm} 10.6112  &   CSO  &   7.8  &  24  &  1.0748  &  0.1829  &  58.1  &  1864  &  38.8  \\
Nov. \hspace{0.000cm} 11.5986  &   CSO  &   8.2  &  30  &  1.0773  &  0.1876  &  57.8  &  1912  &  39.8  \\
Nov. \hspace{0.000cm} 12.6163  &   CSO  &   7.8  &  28  &  1.0800  &  0.1924  &  57.5  &  1962  &  40.9  \\
Nov. \hspace{0.000cm} 13.5404  &  JCMT  &   0.8  &   4  &  1.0826  &  0.1969  &  57.1  &   980  &  20.4  \\
Nov. \hspace{0.000cm} 13.5903  &   CSO  &   7.4  &  25  &  1.0827  &  0.1972  &  57.1  &  2010  &  41.9  \\
Dec. \hspace{0.030cm} 15.6115  &  JCMT  &   0.8  &   4  &  1.2471  &  0.3650  &  37.8  &  1816  &  37.8  \\
\enddata
\tablecomments{IRAM and CSO observed HCN \mbox{\textit{J}(3--2)} and
JCMT observed HCN \mbox{\textit{J}(4--3)}.} \tablenotetext{a}{Time
coverage between the first and last spectrum.}
\tablenotetext{b}{Number of collected spectra.}
\tablenotetext{c}{Heliocentric distance.}
\tablenotetext{d}{Topocentric distance.}
\tablenotetext{e}{Topocentric phase angle.} \tablenotetext{f}{HWHM
of the main beam at comet distance; at the observed frequencies the
HWHM is $4.4\arcsec$ for IRAM, $6.9\arcsec$ for JCMT, and
$14.1\arcsec$ for CSO.} \tablenotetext{g}{Minimum escape time from
the main beam, needed to reach HWHM with the assumed constant
velocity of 0.8~km~s$^{-1}$.}
\end{deluxetable}

\clearpage

\clearpage

\clearpage

\clearpage

\begin{thebibliography}{}

\bibitem[A'Hearn et al.(2005)]{AHe05} A'Hearn, M. F., et al., 2005, Science, 310, 258

\bibitem[A'Hearn et al.(2011)]{AHe11} A'Hearn, M. F., et al., 2011, Science, submitted

\bibitem[Belton \& Drahus(2007)]{Bel07} Belton, M. J. \& Drahus, M., 2007, \baas, 39, 498

\bibitem[Belton et al.(1991)]{Bel91} Belton, M. J. S., Mueller, B. E. A., Julian, W. H., Anderson, A. J., 1991, \icarus, 93, 183

\bibitem[Belton et al.(2011)]{Bel11} Belton, M. J. S, et al., 2011, \icarus, in press

\bibitem[Biver et al.(2002a)]{Biv02a} Biver, N., et al., 2002a, EM\&P, 90, 5

\bibitem[Biver et al.(2002b)]{Biv02b} Biver, N., et al., 2002b, EM\&P, 90, 323

\bibitem[Biver et al.(2010)]{Biv10} Biver, N., Bockel\'ee-Morvan, D., Crovisier, J., Lecacheux, A., Frisk, U., Floren, H.-G., Hjalmarson, A., 2010, CBET, 2524

\bibitem[Bockel\'ee-Morvan et al.(2004)]{Boc04} Bockel\'ee-Morvan, D., Crovisier, J., Mumma, M. J., Weaver, H. A., 2004, in Comets II, eds. M. C. Festou, H. U. Keller, H. A. Weaver (Tucson: Univ. of Arizona Press), 391

\bibitem[Combi et al.(2004)]{Com04} Combi, M. R., Harris, W. M., Smyth, W. H., 2004, in Comets II, eds. M. C. Festou, H. U. Keller, H. A. Weaver (Tucson: Univ. of Arizona Press), 523

\bibitem[Davidsson(2001)]{Dav01} Davidsson, B. J. R., 2001, \icarus, 149, 375

\bibitem[Drahus(2009)]{Dra09} Drahus, M., 2009, \emph{Microwave observations and modeling of the molecular coma in comets}, Ph. D. thesis (Univ. of G\"ottingen)

\bibitem[Drahus \& Waniak(2006)]{Dra06} Drahus, M. \& Waniak, W., 2006, \icarus, 185, 544

\bibitem[Drahus et al.(2010)]{Dra10} Drahus, M., K\"uppers, M., Jarchow, C., Paganini, L., Hartogh, P., Villanueva, G. L., 2010, \aap, 510, A55

\bibitem[Feaga et al.(2007)]{Fea07} Feaga, L. M., A'Hearn, M. F., Sunshine, J. M., Groussin, O., Farnham, T. L., 2007, \icarus, 191, 134

\bibitem[Guti\'errez et al.(2002)]{Gut02} Guti\'errez, P. J., Ortiz, J. L., Rodrigo, R., L\'opez-Moreno, J. J., Jorda, L., 2002, EM\&P, 90, 239

\bibitem[Hartley(1986)]{Har86} Hartley, M., 1986, IAU~Circ. 4197

\bibitem[Jewitt(1992)]{Jew92} Jewitt, D. C., 1992, in Proceedings of the 30th Li\`ege International Astrophysical Colloquium, eds. A. Brahic, J.-C. Gereard, J. Surdej (Li\`ege: Univ. Li\`ege Press), 85

\bibitem[Jewitt(1999)]{Jew99} Jewitt, D., 1999, EM\&P, 79, 35

\bibitem[Lisse et al.(2009)]{Lis09} Lisse, C. M., 2009, PASP, 121, 968

\bibitem[Richardson et al.(2007)]{Ric07} Richardson, J. E., Melosh, H. J, Lisse, C. M., Carcich, B., 2007, \icarus, 190, 357

\bibitem[Samarasinha et al.(1986)]{Sam86} Samarasinha, N. H., A'Hearn, M. F., Hoban, S., Klinglesmith, D. A., 1986, in ESA Proc. of the 20th ESLAB Symp. on the Exploration of Halley's Comet (ESA \mbox{SP-250}, vol.~1), 487

\bibitem[Snodgrass et al.(2011)]{Sno11} Snodgrass, C., Fitzsimmons, A., Lowry, S. C., Weissman, P., 2011, \mnras, in press

\end{thebibliography}
\end{document}